\documentclass[trackchanges, twocolumn]{aastex701}

\usepackage{newtxtext,newtxmath}
\usepackage{graphicx}	% Including figure files
\usepackage{dblfloatfix}
\usepackage{amsmath}	% Advanced maths commands
\usepackage{threeparttable}
\usepackage{subcaption}
\usepackage{caption}
\usepackage{color}
\usepackage[dvipsnames]{xcolor}
\usepackage[normalem]{ulem}
\usepackage{booktabs}
\usepackage{multirow}
\usepackage{mathtools}

\newcommand{\fermi}{\textit{Fermi}-{\rm LAT}}

\newcommand{\gray}{$\gamma$-ray~}
\newcommand{\grays}{$\gamma$-rays~}

\newcommand\HI{\ion{H}{1}}
\newcommand\HII{\ion{H}{2}}

\def\deg{\hbox{$^\circ$}}
\begin{document}

\title{Measurement of  Cosmic-Ray Density in the Spiral Arms toward  Galactic Anticenter}

\author[orcid=0009-0001-8776-549X]{Jia-hao Liu}
\affiliation{School of Astronomy, University of Science and Technology of China, Hefei, Anhui 230026, People's Republic of China}
\email{ljhstpc11@mail.ustc.edu.cn}  

\author[orcid=0000-0002-5965-5576]{Bing Liu}
\affiliation{Purple Mountain Observatory, Chinese Academy of Sciences, Nanjing 210023, People's Republic of China}
\email{liubing@pmo.ac.cn}

\author[orcid=0000-0001-5801-2547]{Rui-zhi Yang}
\altaffiliation{Corresponding author: Rui-zhi Yang}
\affiliation{School of Astronomy, University of Science and Technology of China, Hefei, Anhui 230026, People's Republic of China}
\affiliation{TIANFU Cosmic Ray Research Center, Chengdu, Sichuan 610213, People's Republic of China}
\email[show]{yangrz@ustc.edu.cn}

\begin{abstract}
Using nearly 17 years of \textit{Fermi}-LAT data, we analyzed the diffuse \gray emission associated with different spiral arms of the Milky Way in the ranges $l=105\deg$ to $145\deg$ and $b=-5\deg$ to $5\deg$. The \gray emissions from these spiral arms all exhibited clear pion-bump features, which can be well explained by a power-law spectrum of cosmic-ray protons. The spectral indices corresponding to the three distinct spiral arms—the Local Arm, the Perseus Arm, and the Outer Arm—are found to be approximately $-2.75$, $-2.55$, and $-2.80$, respectively. The energy densities of the cosmic rays above 10 GeV are estimated to be all around $0.2 \rm~eV~cm^{-3}$.
We further compared our results with previous studies, which have established various propagation models predicting that cosmic ray density decreases and spectra soften with increasing Galactocentric radius. We found higher densities and harder spectral indices than predicted in the outer Galaxy, suggesting possible contributions from a larger halo, slower magnetic field decay, or a combination of these and other effects. In particular, the spectral hardening observed in the Perseus Arm may be attributed to locally enhanced cosmic-ray acceleration.

\end{abstract}



\section{Introduction}
Cosmic rays (CRs) constitute a population of relativistic particles, including nuclei from hydrogen to actinides, antiprotons, electrons, and positrons, with energies spanning from 1$~\rm MeV$ to beyond 10$^{21}~\rm eV$. Investigating their origin and distribution has been a cornerstone of modern astrophysics. CRs play a multifaceted role in Galactic ecosystems. They deposit energy into the interstellar medium (ISM), drive chemical processes through ionization, and may influence the dynamics of interstellar gas. A fundamental challenge in this field arises from the nature of CR propagation: as these charged particles diffuse through the turbulent Galactic magnetic field, their trajectories are scrambled, largely erasing directional information about their sources by the time they reach Earth. Consequently, local observations probably provide an unrepresentative sample of the Galactic CR population \citep{amato2018}. This indicates that we can not directly infer the large-scale distribution and the spectral characteristics of CRs in other parts of the Galaxy from the local observations.

Indirect observations via \grays provide a powerful method to overcome the limitations of local measurements and probe CRs throughout the Galaxy \citep{Ackermann_2012b,Ackermann_2012a,Fermi2016,Aharonian_2019}. This technique relies on the fact that CR nuclei interact with interstellar gas to produce neutral pions, which decay into \grays, thereby creating a glow that traces the product of the CR density and the gas density. By analyzing this diffuse \gray emission, one can measure the CR density and spectrum in distant regions, allowing for a Galactic-scale study of CR populations. This technique spans a broad range of spatial scales, from Galaxy-wide distribution down to parsec-scale dense clumps where CR penetration is hindered by local magnetic fields \citep{Yang2023}. Such studies are crucial because the distribution of CRs throughout the Milky Way can provide information about their residence time, propagation mechanisms, and the coupling between sources and the ISM. A key finding from previous observations is that the density of GeV CR nuclei is remarkably uniform across nearby interstellar clouds and the Local Arm, showing variations of less than 30\% (e.g., \cite{Abdo_2009,Abdo_2010}, \cite{Ackermann_2011,Ackermann_2012a}). 
However, this picture becomes more complex on Galactic scales. In the inner Galaxy ($l \approx 30\degr$ and $330\degr$), observations reveal a hard-spectrum \gray component associated with \HII\ regions near the 4-kpc ring, indicating localized CR acceleration by young massive star clusters superimposed on the uniform sea \citep{liu2022_hii}. 
Meanwhile, a significant gradient problem exists in the outer Galaxy: the CR density decreases by only 20-40\% from the Solar Circle to the outer Galaxy, a much shallower decline than expected given the rapid drop in supernova remnant densities \citep{Pineda_2013}. 
This challenges conventional propagation models and suggests the diffusion coefficient is coupled to the distribution of CR sources \citep{evoli2012}. 
%The spectrum of CRs inferred locally from direct measurements also differs from the spectrum that pervades the interstellar medium near the Sun; specifically, the local spectrum observed below a few GeV is heavily modified by solar modulation, requiring demodulation to recover the true interstellar spectrum.
%Former examples include the studies on Westerlund 2 \citep{2018yang}, W43 \citep{w43}, Carina \citep{carina}, G25 \citep{RSGC1},  NGC 3603 \citep{2017yang}, NGC 6618 \cite{m17}, and NGC 2244 \citep{rosette}, 

The LHAASO-KM2A experiment detected significant diffuse \gray emission from the Galactic plane in the energy range of 10$~\rm TeV$ to 1$~\rm PeV$, revealing excesses in both the inner (15$\deg < l < 125\deg$, $|b| < 5\deg$) and outer (125$\deg < l < 235\deg$, $|b| < 5\deg$) Galaxy regions \citep{LHAASO_diffuse}. Specifically, in the outer Galaxy region, which corresponds to the anticenter direction, the measured flux is higher than the prediction from hadronic interactions between locally measured cosmic rays and the interstellar medium by a factor of approximately 2 for energies up to 60$~\rm TeV$. This excess is unexpected under the conventional assumption of uniform cosmic ray intensity, as cosmic ray densities should be lower in the outer Galaxy compared to the inner regions. The discrepancy may arise from several potential origins, including contributions from unresolved populations of \gray emitters, inverse Compton scattering by electrons and positrons injected from pulsars or pulsar wind nebulae, interactions of cosmic rays with material near acceleration sites, or spatial variations in cosmic ray spectra. Further investigation is necessary to determine the exact cause of this situation.

The recent Data Release 1 (DR1) of the Milky Way Imaging Scroll Painting (MWISP) survey \citep{MWISP_DR1} provides an unprecedented set of high-sensitivity, multi-line CO isotopologue data for the northern Galactic plane. This high-quality dataset, characterized by its complete spatial sampling and excellent velocity resolution, significantly enhances our understanding of the outer Galactic disk structure. The work of \cite{du2016} kinematically decomposes the interstellar gas in the outer Galaxy within the Galactic longitude range of $l=100\deg$ to 150$\deg$ into distinct spiral arm components, primarily the Perseus Arm, the Outer Arm, and the New Arm, based on their Local Standard of Rest (LSR) velocity intervals. This kinematic decomposition provides discrete spatial templates of gas distribution for each arm, enabling the possibility of studying the cosmic-ray distribution by using these arm-specific gas templates to analyze \gray data.

In this work, we employ nearly 17 years of \fermi\ data combined with information on the gas distribution in spiral arms, in an attempt to derive the cosmic-ray distribution at varying distances from the Galactic center. In Sec.~\ref{sec:data}, we analyzed \fermi\ data and studied the spatial and spectral distribution of \gray emissions in different spiral arms of the Galaxy. In Sec.~\ref{sec:cr}, we derived the CR spectrum based on the analysis results of Sec.~\ref{sec:data}, investigated how the CR energy densities and spectral indices vary with Galactocentric distance, and discussed the possible contribution of inverse-Compton emission to the observed \gray spectra.
Finally, in Sec.~\ref{sec:dis}, we discussed our results along with previous works and drew our conclusions.

\label{sec:intro}

\section{Fermi-LAT data analysis}
\label{sec:data}
We analyzed data from \fermi\ using the latest Pass 8 data releases, covering the period from August 4, 2008 (MET 239557417) to May 5, 2025 (MET 768172085). The analysis was performed using the Fermitools software package, which was installed and managed via the Conda distribution\footnote{\url{https://github.com/fermi-lat/Fermitools-conda/}}.

We utilize data from the DR1 of the MWISP survey \citep{MWISP_DR1} and the data cube of the \HI\ $\rm{4\pi}$ survey (HI4PI, \cite{HI4PI16}) to obtain the distribution templates of molecular hydrogen (H$_{2}$) and the neutral atomic hydrogen (\HI) for different spiral arms, respectively. Additionally, we employ an all-sky map of dust opacity from the Planck collaboration \citep{2014Planck} to construct the template for the total gas column density using dust distribution, and generate the distribution template for the dark neutral medium (DNM). The detailed calculation procedures are described in  APPENDIX~\ref{sec:gas}.
\subsection{Fitting of Catalog Sources}
\label{sec:fit_catalog}
We first utilized the data in the energy range of 1–500$~\rm GeV$ to perform the spatial analysis. Taking into consideration both the coverage of the gas observational data and the distribution of \gray sources, we ultimately selected a region spanning Galactic longitudes from 105$\deg$ to 145$\deg$ and Galactic latitudes from -5$\deg$ to 5$\deg$ as the region of interest (ROI) for this study. Given the large spatial extent of the ROI, performing a fit as a single integrated domain would involve an excessive number of free parameters. Therefore, we began by modeling other \gray sources in the LAT 12-year Source Catalog (4FGL-DR3, \cite{4FGL}) within three separate ROIs. Each ROI is defined as a $20\deg \times 20\deg$ square, centered at $(l,\; b) = (111\deg,\; 0\deg)$, $(125\deg,\; 0\deg)$, and $(139\deg,\; 0\deg)$, respectively.
%For this work, we defined a region of interest (ROI) within the Galactic plane, spanning 105$\deg$ to 145$\deg$ in Galactic longitude and -5$\deg$ to +5$\deg$ in Galactic latitude. 
The data extraction regions were defined as circles with a radius of 15$\deg$ centered at the same coordinates to ensure they fully contained the subsequent rectangular analysis region.

\begin{figure*}
    \centering
    % 第一行：第一个子图（counts map）
    \begin{subfigure}[t]{0.9\linewidth}
        \centering
        \includegraphics[width=\linewidth]{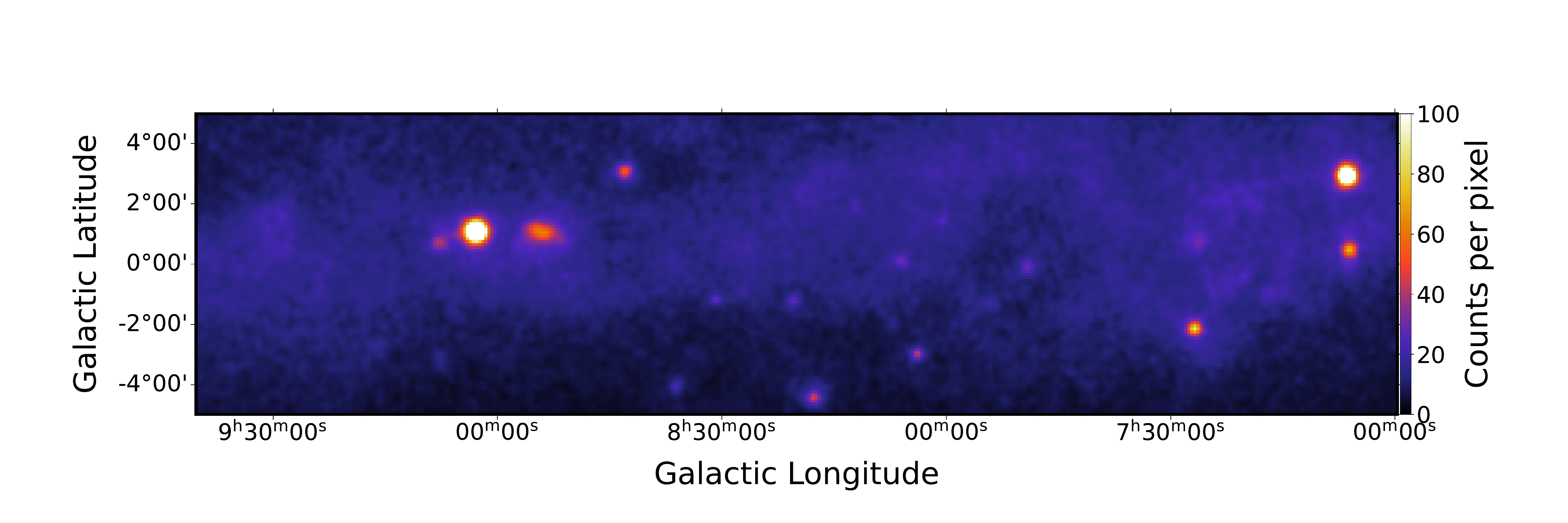}
        \caption{Counts map of \grays in the 1 - 500$~\rm GeV$ energy range.}
        \label{fig:cmap}
    \end{subfigure}
    
    % 第二行：第二个子图（residual significance）
    \begin{subfigure}[t]{0.9\linewidth}
        \centering
        \includegraphics[width=\linewidth]{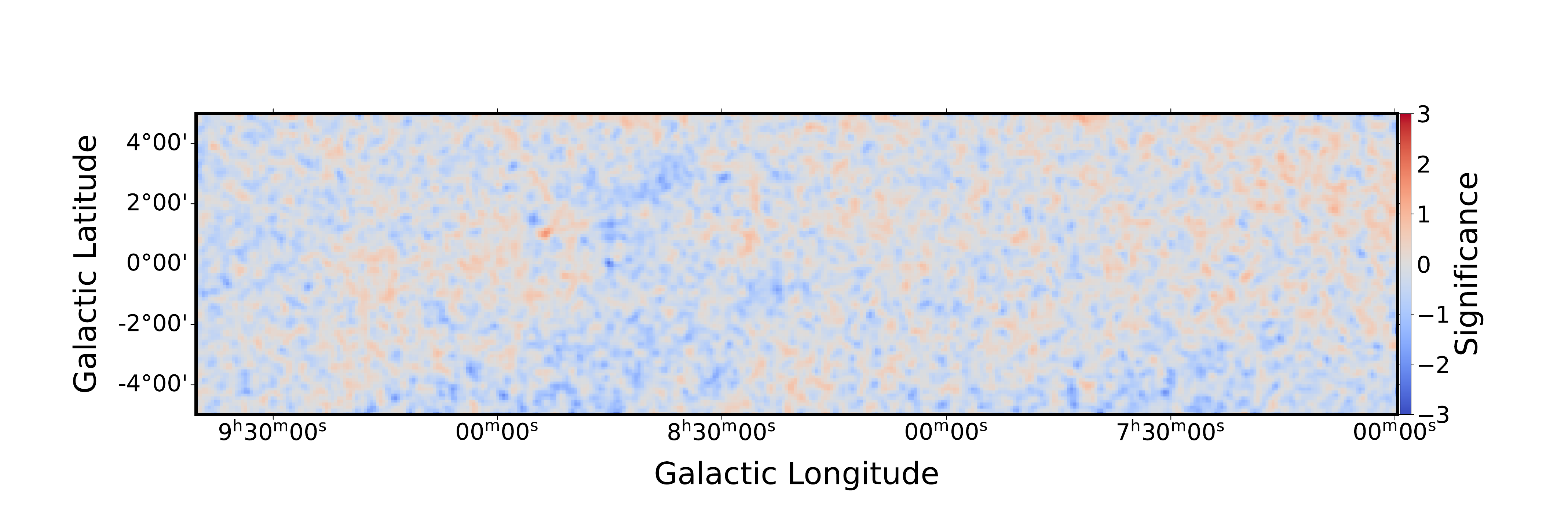}
        \caption{Residual significance map of the 4-part model, corresponding to \gray from 1 to 500$~\rm GeV$. The residual significance is calculated as the difference between the predicted and actual counts, divided by the square root of the latter.}
        \label{fig:res}
    \end{subfigure}
    
    \vspace{0.4cm}  % 行间距
    
    % 第三行：两个子图并排
    \begin{subfigure}[t]{0.48\linewidth}
        \centering
        \includegraphics[width=\linewidth]{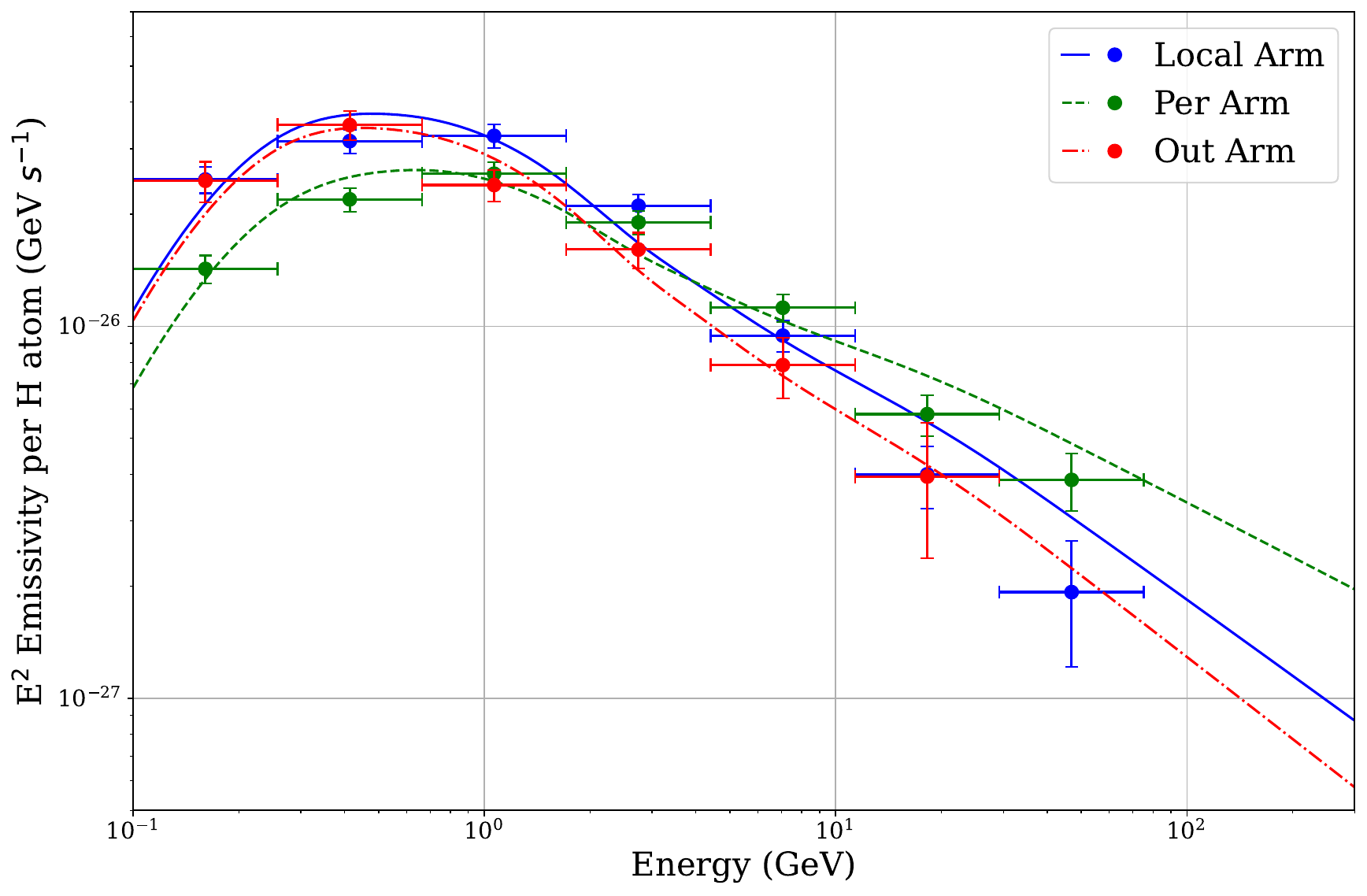}
        \caption{\grays emissivity per hydrogen atom of different spiral arms. The curves represent the \gray emissivities corresponding to the cosmic-ray spectra fitted in Sec. \ref{sec:cr_had}.}
        \label{fig:gamma}
    \end{subfigure}
    \hfill
    \begin{subfigure}[t]{0.48\linewidth}
        \centering
        \includegraphics[width=\linewidth]{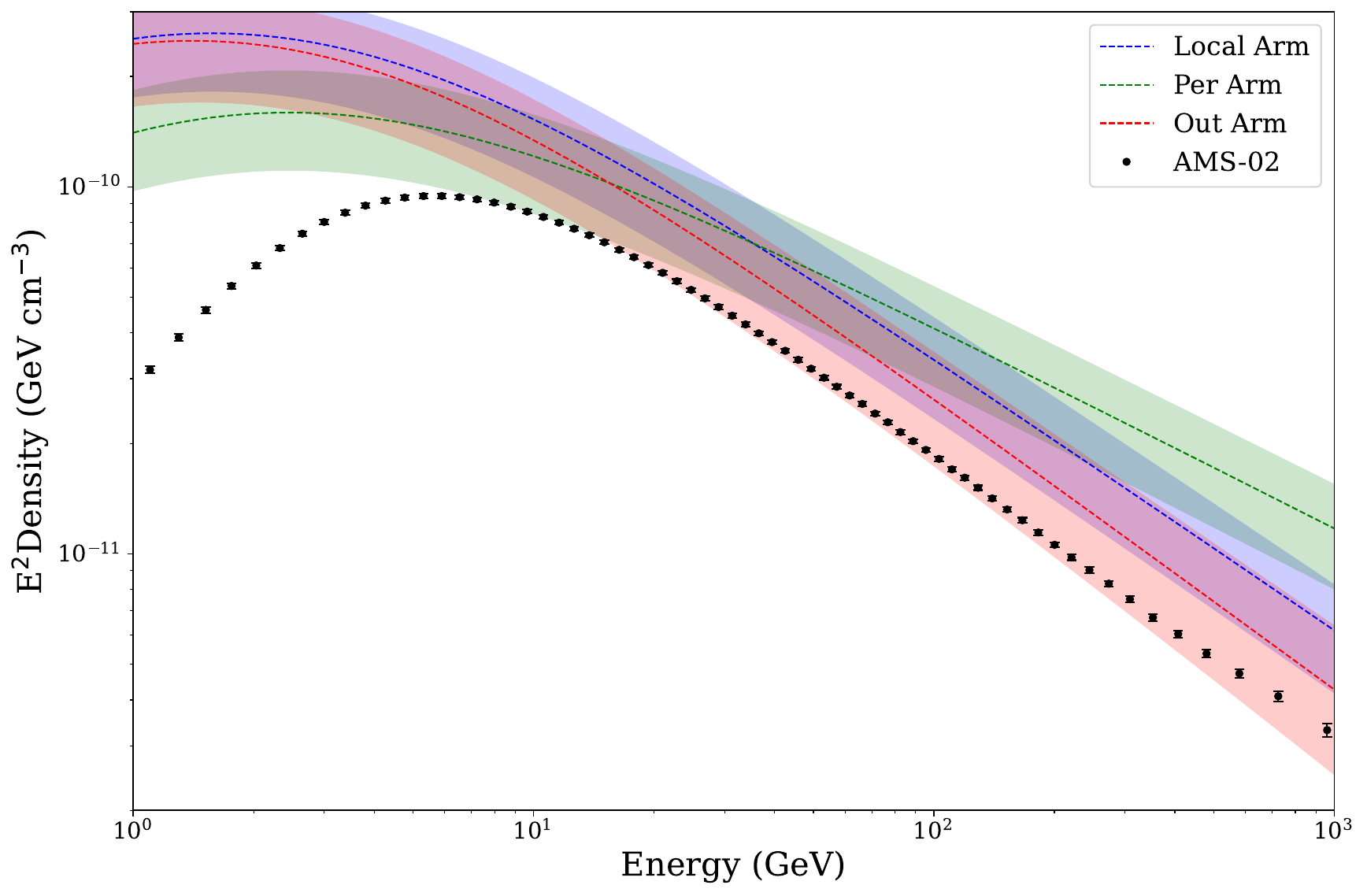}
        \caption{Spectra of CRs in different spiral arms. The shaded areas represent the 1$\sigma$ confidence interval.}
        \label{fig:cr}
    \end{subfigure}
    
    \caption{The results of our \gray spatial analysis (panels a and b), spectral analysis (panel c), and CR spectra fitting (panel d). The black dots in panel d are the AMS-02 observation data acquired from \citet{AMS02_proton}.}
    \label{fig:fermi}
\end{figure*}

The data reduction followed the standard \fermi\ analysis procedures as outlined in the official binned likelihood tutorial\footnote{\url{https://fermi.gsfc.nasa.gov/ssc/data/analysis/scitools/binned_likelihood_tutorial.html}}. We used the {\sl gtselect} to filter the event data, selecting "source" class events (evclass=128) with a maximum zenith angle of 90$\deg$ to minimize contamination from the Earth's limb. The good time intervals were selected using the {\sl gtmktime} with the standard filter expression $\rm (DATA\_QUAL > 0) \&\& (LAT\_CONFIG == 1)$ to ensure data quality. All spatial selections and maps were generated in the Galactic coordinate system. A \gray counts map of the ROI can be seen in Fig.~\ref{fig:cmap}.

The instrument response functions {\it P8R3\_SOURCE\_V3} were used throughout the analysis. A binned likelihood analysis was performed. The source model for the ROI was constructed using the make4FGLxml.py\footnote{\url{https://fermi.gsfc.nasa.gov/ssc/data/analysis/user/make4FGLxml.py}}, which incorporates sources from the catalog. The model includes all 4FGL-DR3 sources located within the ROI and an additional 10$\deg$ buffer, the Galactic diffuse emission model (gll\_iem\_v07.fits), and the isotropic diffuse background template (iso\_P8R3\_SOURCE\_V3\_v1.txt).
In the likelihood fit, the spectral parameters of sources within 10$\deg$ from the center of the ROI were set free. The normalization parameters of both the Galactic and isotropic diffuse components were also left free during the fitting process.
\subsection{Fitting of the Gas Templates}
\label{sec:fit_gas}
We then incorporated templates for different spiral arms and various gas components into our fitting.  
All catalog sources were fixed to the values obtained from the fitting in Sec.~\ref{sec:fit_catalog}. For sources that appeared multiple times in the selected ROIs, we calculated their angular distances to the center of each ROI and adopted the fitting results from the ROI with the smallest distance. Since such sources may significantly affect the diffuse gas flux calculation, we left the parameters of highly significant sources (exceeding $15\sigma$) free. 

Subsequently, we used the gas templates constructed in APPENDIX~\ref{sec:gas} to replace the original Galactic diffuse emission model to fit the \gray emission from different spiral arms. In the first source model, we used only the dust template to fit all diffuse \gray emission. In the second source model, we fitted all gas phases from the three spiral arms and the DNM template separately. We refer to this model as the 7-part model. To determine which model best fits the data, we used the Akaike information criterion (AIC), defined as $-2 \log(\mathcal{L}) + 2k$, where $k$ is the number of free parameters. The values of $-\log(\mathcal{L})$ and the corresponding AIC for each model are listed in Table~\ref{tab:para}. The results show that, compared to using only the dust template, the 7-part model yields a significant improvement in the fit. 

However, during the fitting process, we found that the CO template for the Outer arm did not converge. This is likely because the CO emission in this region is too sparse to be detected significantly, a point that is also evident in Fig.~\ref{fig:out_co}. Therefore, to better fit the \gray emission associated with this gas component, we adopted an assumed $X_{\rm CO}$ of \textbf{$\rm 2.0 \times 10^{20}\ cm^{-2}\ K^{-1}\ km^{-1}\ s$} to derive the H$_2$ column density distribution. By combining this with the \HI\ column density distribution, we obtained the total H atom column density distribution for the Outer arm. This derived H column density map was then used as a new template, replacing the original CO and \HI\ templates for this component. We refer to such templates as gas templates. The model resulting from this replacement is referred to as the 6-part model. In particular, we tested cases with $X_{\rm CO}$ values of $\rm 3.0 \times 10^{20}\ cm^{-2}\ K^{-1}\ km^{-1}\ s$ and $\rm 4.0 \times 10^{20}\ cm^{-2}\ K^{-1}\ km^{-1}\ s$ and found that they had almost no effect on the fitting results, especially for the Local and Perseus arms.
 
For the diffuse gas in the different spiral arms, the \HI\ column densities are generally modest, so the effects of \HI\ optical-depth saturation and self-absorption are expected to be small. 
Under these conditions, the \HI\ column densities can be regarded as reliable estimates of the atomic gas content. The conversion from CO intensity to H$_2$ column density, however, depends on the uncertain $X_{\rm CO}$ factor, which is known to vary with environmental conditions. For the gamma-ray emission associated with the CO and \HI\ templates from the same spiral arm, the cosmic-ray environment can be assumed to be the same, so the gamma-ray fluxes should be proportional to the respective column densities of molecular and atomic gas. We therefore use the well-determined \HI\ gas as a reference to calibrate $X_{\rm CO}$. Consequently, after converting to total H column density, the gamma-ray emission associated with the CO and \HI\ templates from the same spiral arm should have the same proportionality. Based on this premise, we derived the actual $X_{\rm CO}$ values for the Local and Perseus arms, finding both to be approximately $8 \times 10^{19}\ \rm cm^{-2}\ K^{-1}\ km^{-1}\ s$.
%Under these conditions, the \HI\ column densities can be regarded as reliable estimates of the atomic gas content and are therefore used as the reference for calibrating the molecular gas traced by CO emission.
%Consequently, after converting to total H column density, the \gray emission associated with the CO and \HI\ templates from the same spiral arm should have the same proportionality. Based on this premise, we derived the actual $X_{\rm CO}$ values for the Local and Perseus arms, finding both to be approximately $\rm 8 \times 10^{19}\ cm^{-2}\ K^{-1}\ km^{-1}\ s$. 
Similar to the Outer arm, we calculated the total H column density distributions for these two arms based on their derived $X_{\rm CO}$ values. These new H column density maps were then used as new templates, replacing the original CO and \HI\ templates for the respective gas components. The resulting model is referred to as the 4-part model. The fitting results for the different source models are summarized in Table~\ref{tab:para}. It can be seen that, compared to the dust model which treats all the gas in the region as a single component, the model dividing the gas into different spiral arms yields a significant improvement in the fit. On the other hand, after merging the CO and \HI\ templates of the different spiral arms, the fitting performance remains at a reasonably good level, indicating that our method for calculating the total \gray spectra corresponding to the gas in each spiral arm is feasible. In addition, we also present the residual significance map of the 4-part model in the 1 to 500$\rm~GeV$ energy range (Fig. \ref{fig:res}), which shows no significant residuals within the ROI. Therefore, we applied this 4-part model to the later spectral analysis.

\begin{table}
\caption{Fitting results of different source models. Note that $\Delta\left(-\log\left(\mathcal{L}\right)\right)$ and $\Delta(\mathrm{AIC})$ refer to the changes relative to the plain dust model.}  
\label{tab:para}       
\centering                        
\begin{tabular}{cccc}       
\hline\hline     
Model&$\Delta\left(-\log\left(\mathcal{L}\right)\right)$&k&$\Delta($AIC$)$\\
\hline
%dust&372563&32&745190\\
dust&-&32&-\\
%7-part&372398&44&744884\\
7-part&-165&44&-306\\
%6-part&372416&42&744916\\
6-part&-147&42&-274\\
%4-part&372425&38&744926\\
4-part&-138&38&-264\\
\hline
\end{tabular}
\end{table}

\subsection{Spectral analysis for different spiral arms}
\label{sec:spectra}
To study the \gray spectrum corresponding to each spiral arm, we utilized data from $100\rm~MeV$ to $500\rm ~GeV$ and divided it into 9 energy bins evenly distributed in logarithmic space. For the data from each energy bin, we conducted likelihood analysis with the 4-part model in Sec.~\ref{sec:fit_gas} and derived the total flux of each template. Systematic uncertainties were also estimated using the method described on the Fermi official website\footnote{\url{https://fermi.gsfc.nasa.gov/ssc/data/analysis/scitools/Aeff_Systematics.html}}. The derived \gray spectra normalized to \gray emissivity per hydrogen atom are shown in Fig.~\ref{fig:gamma}. The error bars represent the combined statistical and systematic uncertainties. Uncertainties arising from the different adopted values of $X_{\rm CO}$ for the Outer arm were also taken into account. We note that the spectra of all templates show clear hints of a low-energy break in the first energy bin, a feature consistent with the pion-bump expected in the hadronic scenario of \gray production. Please note that the DNM template is retained in the likelihood analysis because previous studies \citep[e.g.,][]{Acero2016,Remy2017,Youssef2024} have shown that it represents a physically distinct gas component that is not fully traced by conventional gas surveys. Including this component improves the modeling of the interstellar gas distribution and helps account for residual \gray emission associated with otherwise untraced gas. However, its contribution in the present region is relatively weak compared to that of the spiral-arm gas components and does not play a significant role in the scientific conclusions of this work, so we do not further discuss the DNM component in the rest part of this article.

\section{CR content in different spiral arms}
\label{sec:cr}
The distinct pion-decay signature suggests that the \gray emission associated with the spiral arms originates predominantly from proton-proton interactions between CR protons and the interstellar gas. However, a non-negligible contribution from leptonic processes, in particular inverse-Compton (IC) scattering of CR electrons off the interstellar radiation field (ISRF), cannot be excluded. To explore the origin of the observed \gray emission, we therefore investigate both hadronic and leptonic scenarios separately.

\subsection{Hadronic Scenario}
\label{sec:cr_had}
In the hadronic scenario, we investigate the CR content using the \gray data presented in Sec.~\ref {sec:data} and the gas distributions derived in APPENDIX~\ref{sec:gas}. We assume that the cosmic-ray protons follow a momentum distribution function: $F = f_i p^{\Gamma}$, where $F$ is the proton flux density in units of $\mathrm{cm}^{-2}\,\mathrm{s}^{-1}\,\mathrm{GeV}^{-1}$, $p$ is the proton momentum in GeV/c, and $\Gamma$ is the spectral index. Under this assumption, we employed the \gray production cross section from \citet{2014pp} to perform a likelihood fit to the \gray data. The gas masses used to derive the absolute CR fluxes were calculated based on our calculated gas distributions for different spiral arms ( APPENDIX~\ref{sec:gas}), noting that for a resulting gamma-ray flux derived in our analysis, the CR flux is inversely proportional to the gas mass. The derived spectral indices $\Gamma$ of the CR spectrum from different spiral arms can be seen in Tab.~\ref{tab:index}. The corresponding \gray spectra (scaled to \gray emissivity per hydrogen atom) are shown in different lines in Fig.~\ref{fig:gamma}. The derived CR spectra are shown in Fig.~\ref{fig:cr}. The locally measured data from AMS-02 are also plotted in Fig.~\ref{fig:cr}, indicated by the black data points. The deviation between the AMS-02 data and the model predictions at lower energies ($\lesssim 10$~GeV) is attributed to heliospheric modulation, which suppresses the low-energy cosmic-ray flux through diffusion, convection, and adiabatic energy losses as particles propagate through the solar wind \citep{Potgieter2013}.
\begin{table}
\caption{Range of distances from the GC ($d_{\rm{GC}}$) to each spiral arm, along with the corresponding spectral indices ($\Gamma$) and energy densities ($U_{\rm CR}$) of CRs over 10$~\rm GeV$.} 
\label{tab:index}       
\centering                        
\begin{tabular}{cccc}       
\hline\hline     
&Local Arm&Perseus Arm&Outer Arm\\
\hline
$d_{\rm{GC}}$ (kpc)&8.0 - 9.0&8.5 - 10.5&9.0 - 12.8\\
$\Gamma$&-2.75$\pm$0.03&-2.55$\pm$0.02&-2.80$\pm$0.07 \\
$U_{\rm CR}$ ($\rm eV~ cm^{-3}$)&0.23$\pm$0.07&0.25$\pm$0.07&0.19$\pm$0.06 \\
\hline
\end{tabular}
\end{table}

We compared the spectral indices in different spiral arms and thus at various distances from the Galactic Center ($d_{\rm{GC}}$). Based on Figure 1 from the work of \cite{Reid2019} and using the Galactic longitude range of the ROI in this study, with the widths enclosing 90\% of the sources provided in the figure adopted as the spiral arm widths, the range of $d_{\rm{GC}}$ for each spiral arm was estimated and presented in Table~\ref{tab:index}. Specifically, for the $d_{\rm{GC}}$ corresponding to the locally measured cosmic rays, the value of 8.15$\pm0.15~\rm kpc$ from the work of \cite{Reid2019} was used. For the local CR index, we adopted the proton spectral index measured by AMS-02 \citep{AMS02_proton}. For the rigidity range below the spectral hardening break at $R_0 \approx 330 \rm~GV$ (corresponding to $\sim 100~\rm GeV$ in kinetic energy), the spectral index is 2.85 $\pm$ 0.01.
%we adopted 2.772 $\pm$ 0.002 below 0.48$~\rm TeV$, which was obtained from a smoothly broken power-law fit to the DAMPE data \citep{DAMPE_CR}.
\begin{figure*}
    \begin{subfigure}[t]{0.45\linewidth}
        \centering
        \includegraphics[width=\linewidth]{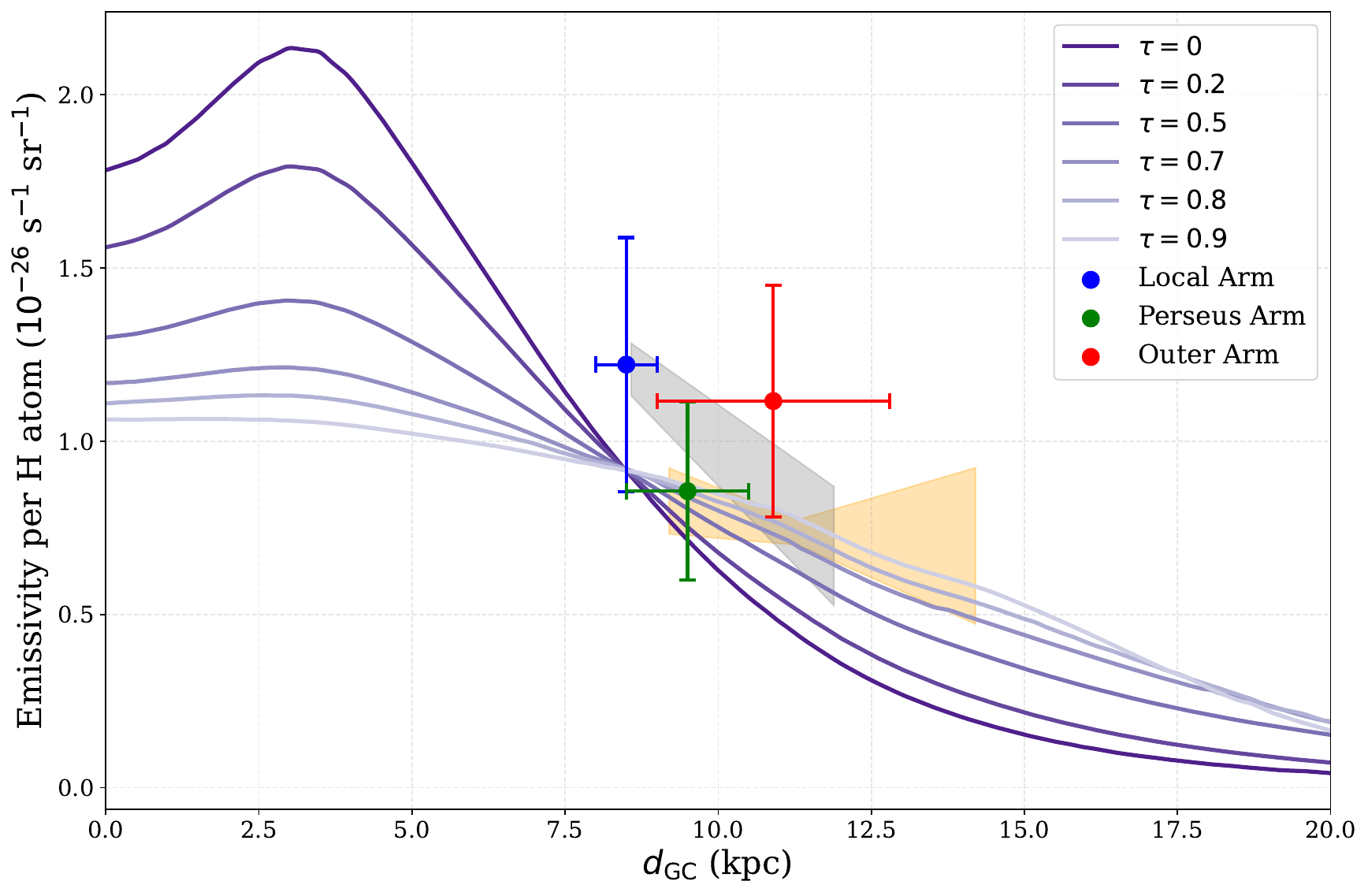}
        \caption{Integrated \gray emissivity in different spiral arms and at different $d_{\rm GC}$. The curves are derived from Fig.~1 in the work of \citet{evoli2012}. Here, the curves from top to bottom correspond to models with increasing values of the diffusion inhomogeneity parameter $\tau=0$, $0.2$, $0.5$, $0.7$, $0.8$, and $0.9$, respectively. The orange and grey shaded regions show observational constraints from \fermi\ data presented by \citet{Ackermann_2011} and \citet{Abdo_2010}, respectively.}
        \label{fig:emi_gradient}
    \end{subfigure}
    \hfill
    \begin{subfigure}[t]{0.45\linewidth}
        \centering
        \includegraphics[width=\linewidth]{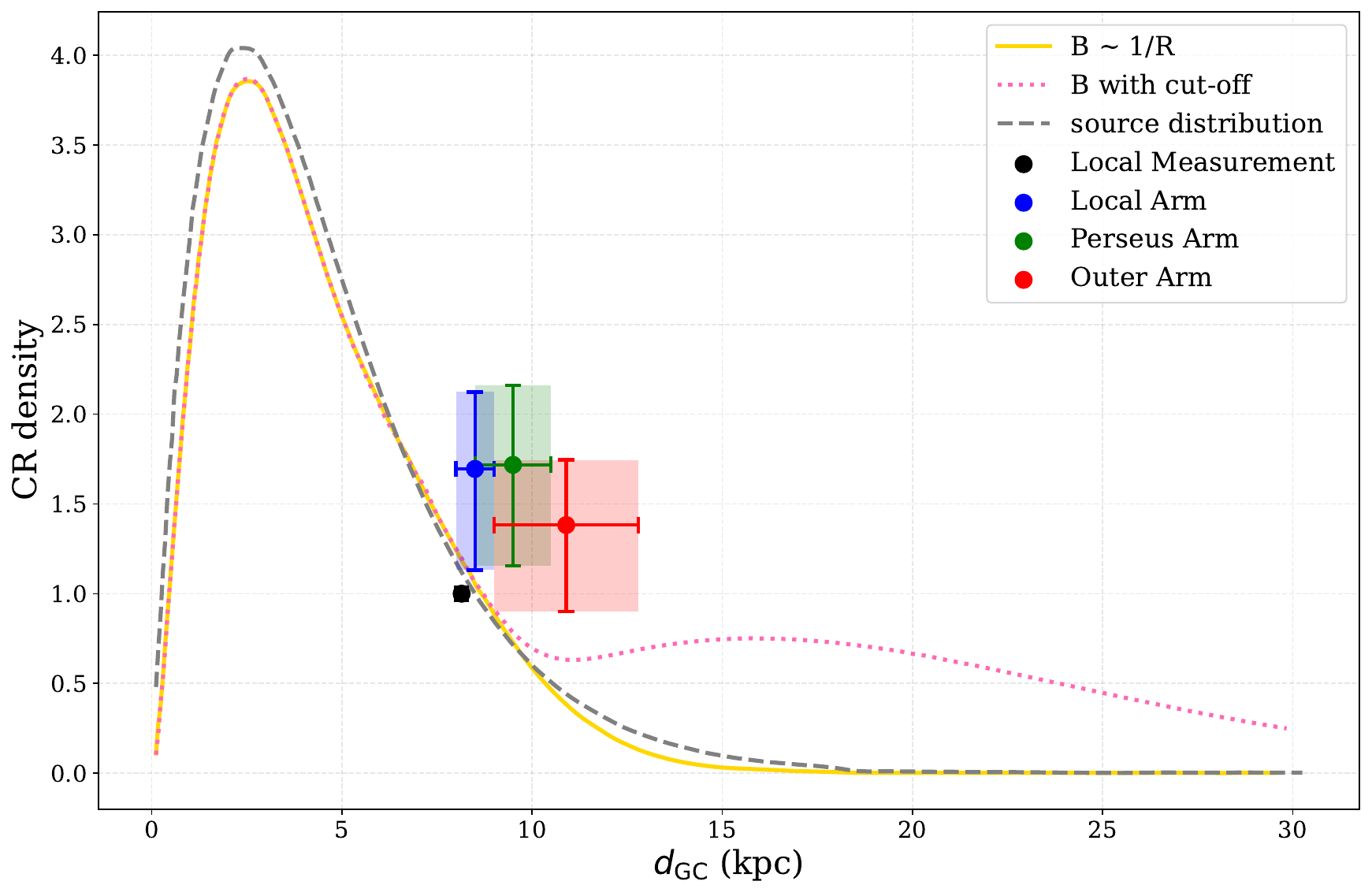}
        \caption{Densities of CRs over 20$~\rm GeV$ in different spiral arms and at different $d_{\rm GC}$. The curves are derived from Fig.1 in \citet{Recchia2016}. The gray dashed curve shows the source distribution model, the yellow solid curve shows $B(R) \propto R^{-1}$, and the pink dashed curve shows the exponential cutoff model. Please note that this figure is scaled such that the local measurement is set to 1, which is consistent with \citet{Recchia2016}.}
        \label{fig:cr_density}
    \end{subfigure}

    % 第三行：单独一个子图
    \centering
    \begin{subfigure}[t]{0.45\linewidth}
        \centering
        \includegraphics[width=\linewidth]{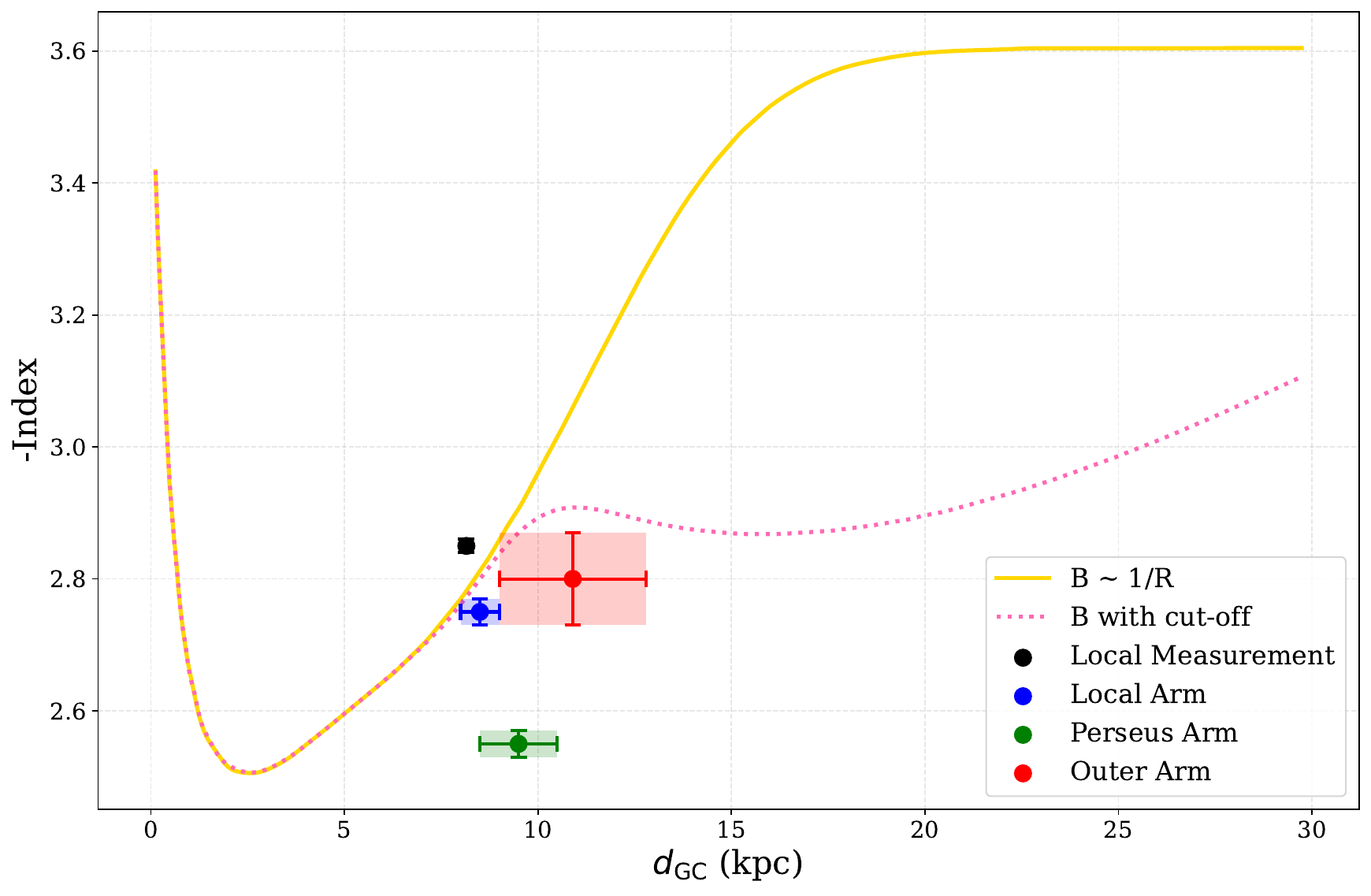}
        \caption{CR spectral indices in different spiral arms and at different $d_{\rm GC}$. The curves are derived from Fig.2 in \citet{Recchia2016}. The yellow solid curve corresponds to $B(R) \propto R^{-1}$, and the pink dashed curve corresponds to the exponential cutoff model.}
        \label{fig:cr_index}
    \end{subfigure}
    
    \caption{CR properties in different spiral arms and comparisons with previous works. The blue, green, and red data points in all subfigures correspond to the results for the Local Arm, Perseus Arm, and Outer Arm obtained in this work, respectively. The black data points represent measurements of local cosmic rays by AMS-02 \citep{AMS02_proton}.}
    \label{fig:cr_properties}
\end{figure*}

Similarly, the cosmic-ray energy density $U_{\rm CR}$ above 10$~\rm GeV$ for different spiral arms was calculated by integration, and the result for the locally measured cosmic rays by AMS-02 \citep{AMS02_proton} was obtained via interpolation. The results for different spiral arms are also shown in Table~\ref{tab:index}.
\subsection{Leptonic Scenario}
\label{sec:cr_lep}
To estimate the IC emission from the selected region, we assume that the cosmic-ray electron density is uniform and equal to the local value. The IC flux is calculated by integrating the line-of-sight contributions over the entire volume.

\begin{figure}
    \centering
    \includegraphics[width=0.9\linewidth]{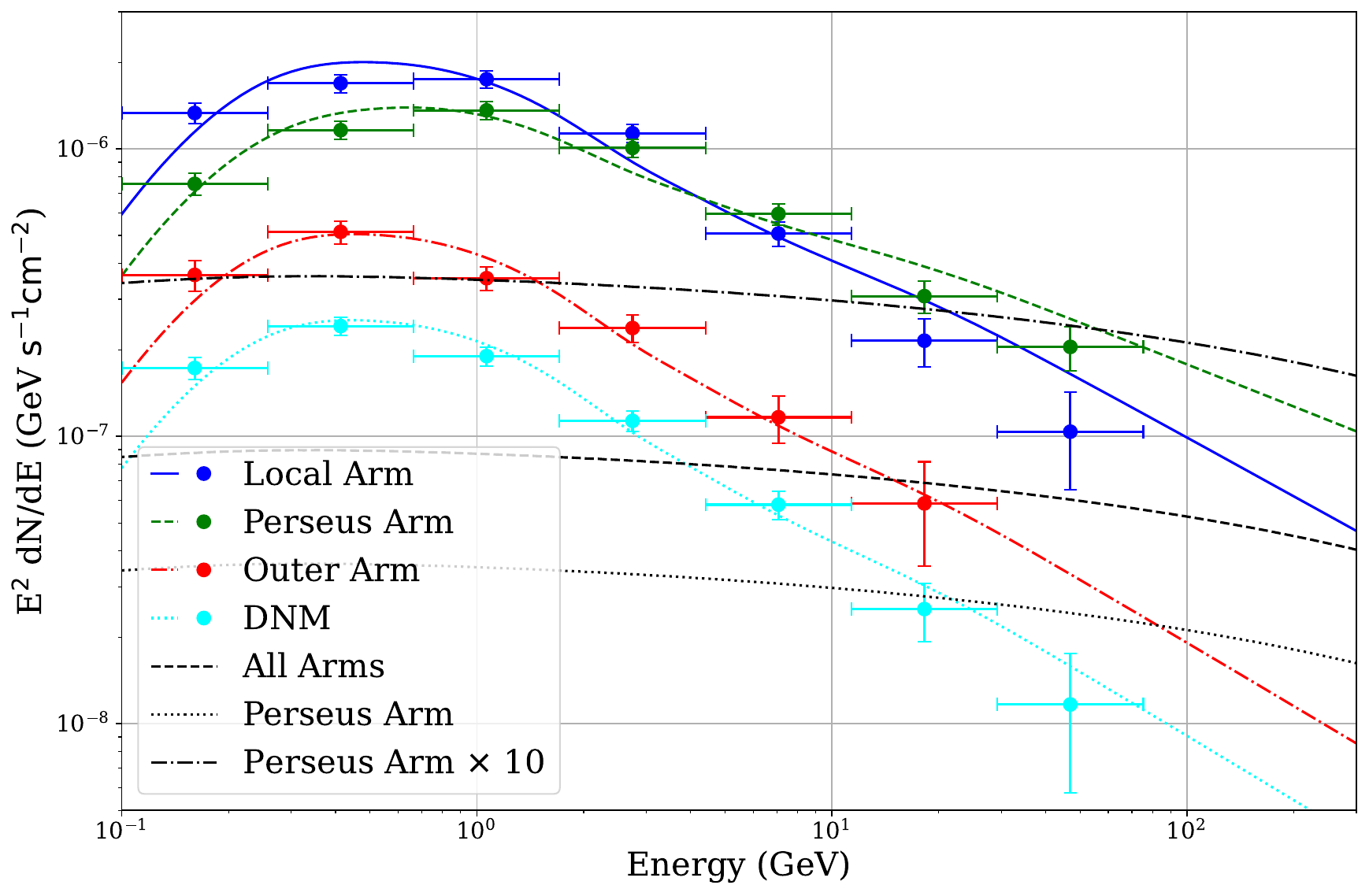}
    \caption{The \gray spectra of different spiral arms. The colored curves represent the \gray spectra corresponding to the cosmic-ray spectra fitted in Sec. \ref{sec:cr_had}. The black curves represent the IC contribution to \gray spectra calculated in Sec. \ref{sec:cr_lep}. }
    \label{fig:gammaflux}
\end{figure}

For the ISRF, we use the Milky Way radiation-field model of \citet{Popescu2017}. The optical, infrared, and CMB components are adopted as the seed photon field in \texttt{naima} \citep{naima}, and the resulting IC spectrum is scaled by the volume integral to obtain the total emission.
We perform the calculation for two spatial ranges: all spiral arms region, corresponding to an integration over $d_{\rm GC}=8.0$--$12.8~\mathrm{kpc}$, and the Perseus arm region, corresponding to $d_{\rm GC}=8.5$--$10.5~\mathrm{kpc}$. These spatial restrictions are adopted to estimate the localized IC contribution from the respective regions, under the assumption that a uniform CR electron density fills the Galaxy. The resulting \gray spectra are shown in Fig. \ref{fig:gammaflux}.

As shown in Fig. \ref{fig:gammaflux}, under the assumption of a locally uniform electron density, the IC emission integrated over the full spiral-arm region can contribute approximately 20\%--30\% of the observed \gray flux above $\sim10~\mathrm{GeV}$ in the Perseus-arm spectrum. Such a contribution may partially account for the apparent hardening at the higher energies. In contrast, when the integration is restricted to the Perseus-arm region alone, the predicted IC emission becomes negligible. 

CR electrons are unlikely to be distributed uniformly throughout the Galaxy. 
We adopt the Galactic CR electron distribution predicted by GALPROP\footnote{\url{http://galprop.stanford.edu/webrun/}} \citep{galprop}, using 
the model parameters from the grid of \citet{Ackermann_2012b}: $z_h=4$~kpc, 
$R_h=20$~kpc, $T_S=10^5$~K (optically thin limit), $E(B-V)$ magnitude cut 
of 2~mag, and Lorimer pulsar distribution for the CR source. This parameter 
choice is comparable to that of \citet{strong11}. Compared to the uniform 
CR electron distribution assumption, we find that the IC contribution would 
decrease by a further $\sim$30\%--40\%. We note that this estimate serves only 
as an approximate reference.

For the IC component to fully account for the high-energy emission of the Perseus arm, a combination of an enhanced average CR electron density and a higher mean ISRF energy density within the integration volume would be required, such that the resulting IC emissivity is approximately an order of magnitude larger than that predicted using the local cosmic-ray electron spectrum and the adopted ISRF model. The corresponding model prediction is shown by the dot-dashed curve in Fig. \ref{fig:gammaflux}. If such an IC contribution is present, the hadronic CR spectrum required to fit the observed \gray emission would be softer than that inferred under the pure hadronic scenario discussed 
in Sec.~\ref{sec:cr_had}.

\section{Discussion and conclusion}
\label{sec:dis}
In this paper, we analyzed almost 17 years of \fermi\ data towards the Galactic anticenter region. Diffuse gas distributions obtained through different methods are employed as templates, in place of the original Galactic diffuse emission model from \fermi, and are validated to provide a good fit to the diffuse \gray emission within the ROI. We found that, compared to the dust template which treats the diffuse gas as a single entity for fitting, significantly better results were obtained by separately fitting the templates corresponding to different spiral arms. This indicates that the cosmic rays in different spiral arms exhibit intrinsic differences. In the obtained \gray energy spectra of the different spiral arms, a pronounced bump feature was identified in each case, offering strong support for the theory that the diffuse \gray radiation originates from collisions between cosmic-ray protons and diffuse gas. 

The \gray SEDs for every arm can be well reproduced using power-law energy spectra for the cosmic-ray protons. By utilizing the fitted energy spectra of cosmic rays from different spiral arms and the local measurements, a comparison was made of the CR spectral index at different distances to the Galactic center. The measured spectral indices for the Local Arm (about -2.75) and the Outer Arm (about -2.80) are broadly consistent with local values (about -2.85). In contrast, the spectrum derived for the Perseus Arm (with an index of approximately $-2.55$) is significantly harder. This spectral hardening may be associated with the presence of superbubbles in this spiral arm. 
Observations of the Cygnus Bubble by LHAASO \citep{LHAASO_cygnus} demonstrate that superbubbles facilitate continuous CR injection via stellar winds from massive star clusters, producing a characteristic CR density profile $n_{\rm CR} \propto 1/r$ \citep{yang2022}. The strongly suppressed diffusion within these bubbles preserves the hard injection spectrum, resulting in a hard \gray spectrum \citep{yang2025}. Similar hard-spectrum emission and continuous injection signatures have been observed in W43 \citep{LHAASO_W43}, the Orion-Eridanus superbubble \citep{Morlino2021}, and near the young clusters Danks~1 and Danks~2 \citep{liu2024}. These findings suggest that stellar winds in superbubbles, coupled with effective local confinement, constitute a key mechanism for sustaining hard-spectrum CRs.
In the context of our study, we note two superbubbles with potential spatial associations: the massive ``Giant Oval'' \citep{Chen2025} in the Perseus Arm (spanning $l \sim 100\deg$ to $150\deg$ at $\sim 2.1$--$2.4~\rm kpc$) and the W4 superbubble \citep{W4}, which also lies within the Perseus Arm and our region of interest. Future work should investigate the potential role of these objects in CR acceleration, e.g., by analyzing individual sources such as the W4 superbubble separately.
Other potential CR accelerators in the Perseus arm include supernova remnants, such as CTB109\citep{CTB109,CTB109_dis}; pulsar wind nebulae, such as 3C 58 \citep{3C58,3C58_dis}; massive star-forming regions, such as W3 \citep{W3,W3_dis}; and young stellar clusters or OB associations, such as NGC7790\citep{NGC7790}. These objects may contribute to the CRs in Perseus arm through shock acceleration, pulsar winds, or collective stellar-wind activity, acting as accelerators of both hadronic and leptonic cosmic rays. 
The CR electrons accelerated in these objects may contribute an additional IC component to the observed \gray emission, hardening the total spectrum at higher energies. This would imply that the underlying hadronic CR spectrum could be softer than that inferred under a pure hadronic scenario. However, for the IC component to fully account for the high-energy part of the \gray emission, the combination of the average CR electron density and the mean ISRF energy density within the relevant volume would need to be approximately an order of magnitude larger than that inferred from the 
local cosmic-ray electron spectrum and the adopted ISRF model. Given the presence of localized CR accelerators in the Perseus arm, such an enhancement of the electron density above the Galactic average is plausible, and the observed hardening may reflect a combination of a moderately enhanced leptonic component and a hadronic spectrum with a 
lower spectral index.
Alternatively, the spectral difference may arise from a varied distribution 
of different CR accelerator populations across the spiral arms.

Regarding the average CR energy density, a trend slightly higher than the local measurement is observed across the spiral arms. This observed trend appears inconsistent with the intuitive expectation of a lower density of accelerators in the anti-Galactic-center direction, yet it aligns with a previously reported excess of diffuse \gray emission in that direction by LHAASO \citep{LHAASO_diffuse}. One possible explanation is that the solar system resides within a local underdensity, or ``void,'' implying the locally measured CR density is lower than the average for the Local Arm. From Fig. 1 in the work of \cite{Elia2022}, it can be seen that the star formation rate in the vicinity of our Solar System is relatively low, both compared to its surrounding regions and to the rates along the same spiral arm. This may be one reason why the local cosmic-ray density is lower. More definitive conclusions require further investigation.

In the context of Galactic CR propagation, different approaches have been proposed to model the radial gradient. The foundational study by \cite{strong1998} constrained the halo height $z_h$ and the diffusion coefficient using secondary-to-primary ratios, assuming spatially uniform transport properties. Subsequently, \cite{evoli2012} introduced a phenomenological model of inhomogeneous diffusion, demonstrating that correlating the perpendicular diffusion coefficient with the source density as $D_\perp \propto Q(R)^\tau$ (with $\tau \approx 0.7$--$0.9$) can simultaneously resolve the gradient and anisotropy problems. In contrast, \cite{Recchia2016} developed a non-linear model where transport is governed by self-generated turbulence, showing that the radial profile of the magnetic field $B_0(R)$ and the competition between advection and diffusion are the critical parameters determining the CR density distribution and the gradual spectral softening from $\Gamma \sim 2.6$ in the inner Galaxy to $\Gamma \sim 2.9$ at larger radii.

We compare our results with the \gray emissivity profile shown in Fig.~1 of \cite{evoli2012} (see Fig.~\ref{fig:emi_gradient}), as well as the CR density and spectral index variations presented in Fig.~1 and Fig.~2 of \cite{Recchia2016} (see Fig.~\ref{fig:cr_density} and Fig.~\ref{fig:cr_index}). Neither the simple model with $B(R) \propto R^{-1}$ nor the model incorporating an exponential cutoff in the magnetic field at $R \gtrsim 10$~kpc provides a fully satisfactory fit to our data: while the latter improves the gradient description compared to the former, both models underestimate the CR density in the outer Galaxy, where we find systematically higher densities in the Outer Arm and Perseus Arm, along with harder spectral indices than predicted. This discrepancy may indicate that the halo height is larger than the fixed value $H=4~\rm kpc$ assumed by \cite{Recchia2016}, as suggested by \cite{strong1998}. Alternatively, the effective diffusion coefficient in the outer disk may be smaller than expected, corresponding to a larger $\tau$ or additional confinement mechanisms beyond the simple $D_\perp \propto Q(R)^\tau$ prescription, or the magnetic field strength may decay more slowly with Galactocentric radius than the $1/R$ or exponential dependence assumed, effectively reducing the diffusion coefficient $D \propto B_0^4/Q^2$ and enhancing CR trapping at large radii. 
In addition, we note that recent three-dimensional cosmic-ray transport studies (e.g., \cite{Johannesson2018, Thaler2023}) suggest that part of the discrepancy between our results and the axisymmetric models of \cite{evoli2012, Recchia2016} may naturally arise from spiral-arm source distributions and localised recent accelerators. These models demonstrate that non-axisymmetric source geometry can lead to enhanced CR densities and systematically harder spectra in spiral arms, without requiring global changes in diffusion or halo parameters. Therefore, the deviations observed in the Perseus and Outer Arms may partly reflect such geometric and source-distribution effects, in addition to possible modifications of large-scale transport properties.
These considerations also apply to the leptonic scenario discussed in Sec.~\ref{sec:cr_lep}. The GALPROP model used to estimate the IC 
contribution is axisymmetric and does not account for localized sources 
such as those present in the Perseus arm. The same three-dimensional 
CR transport studies that predict enhanced hadronic CR densities in 
spiral arms \citep{Johannesson2018, Thaler2023} imply corresponding 
enhancements of the local CR electron density. Such non-axisymmetric structure would increase the IC emissivity above the smooth GALPROP prediction, supporting the possible presence of a non-negligible IC contribution from the Perseus arm.
We note that these possibilities are not exhaustive, and the observed features could arise from a combination of multiple factors; distinguishing among these scenarios will require additional observational data and more detailed theoretical modeling.

\begin{acknowledgments}
R.-z. Y. is supported by the NSFC under grant 12588101, 12393854, and by the natural science funding of Sichuan Province under grant 2025ZNSFSC0065. J.-h. L. is supported by the NSFC under grant 125B2059. Rui-zhi Yang gratefully acknowledges the support of Cyrus Chun Ying Tang Foundations and of the studio of Academician Zhao Zhengguo, Deep Space Exploration Laboratory. B. L. is
supported by the Natural Science Foundation for General
Program of Jiangsu Province of China under grant NO.
BK20252108.
\end{acknowledgments}

\appendix

\section{Gas templates}
\label{sec:gas}
\begin{figure*}[p]  % 使用 [p] 允许单独一页
    \centering
    % 第一行：Local arm
    \begin{subfigure}[t]{0.48\linewidth}
        \centering
        \includegraphics[width=\linewidth]{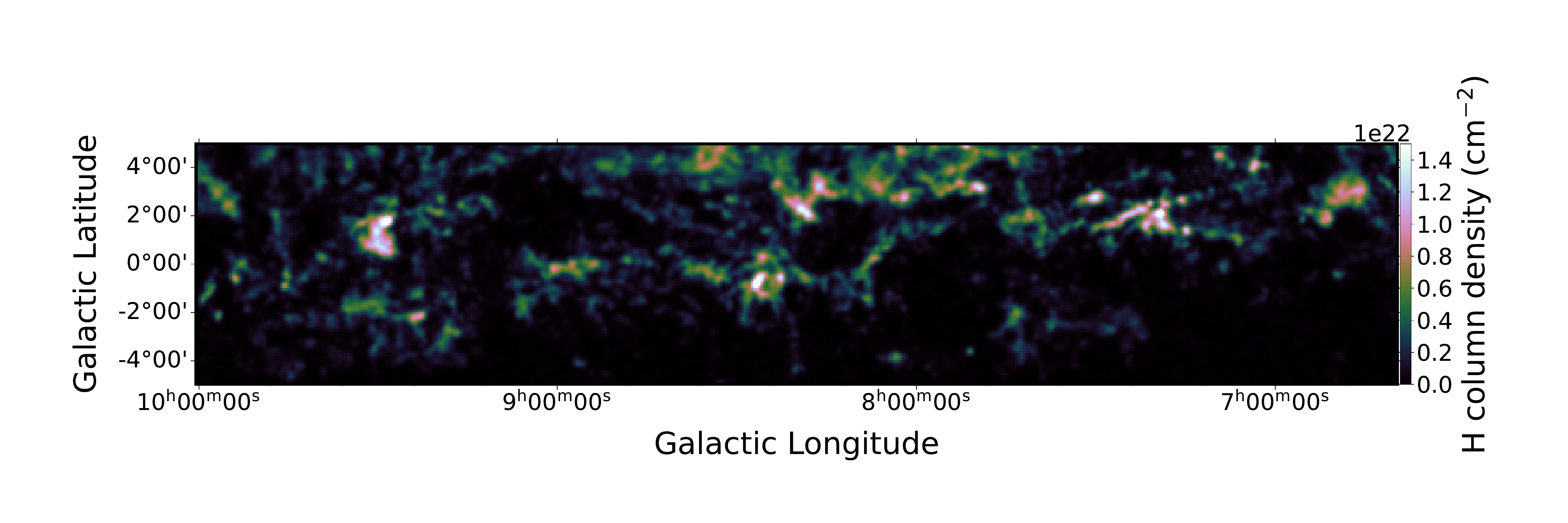}
        \caption{Local arm H$_{2}$}
        \label{fig:loc_co}
    \end{subfigure}
    \hfill
    \begin{subfigure}[t]{0.48\linewidth}
        \centering
        \includegraphics[width=\linewidth]{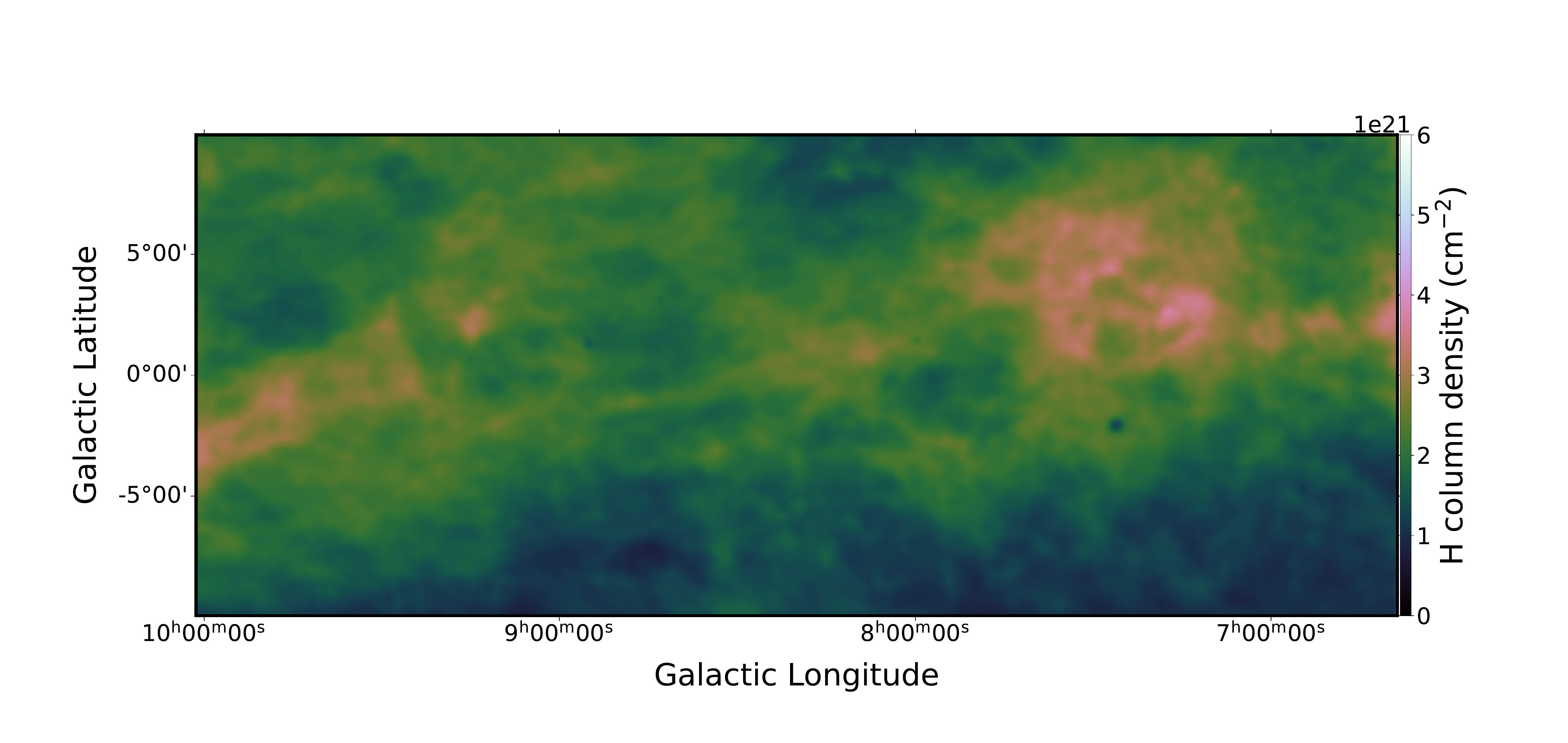}
        \caption{Local arm \HI}
        \label{fig:loc_hi}
    \end{subfigure}
    
    \vspace{0.3\baselineskip}
    
    % 第二行：Perseus arm
    \begin{subfigure}[t]{0.48\linewidth}
        \centering
        \includegraphics[width=\linewidth]{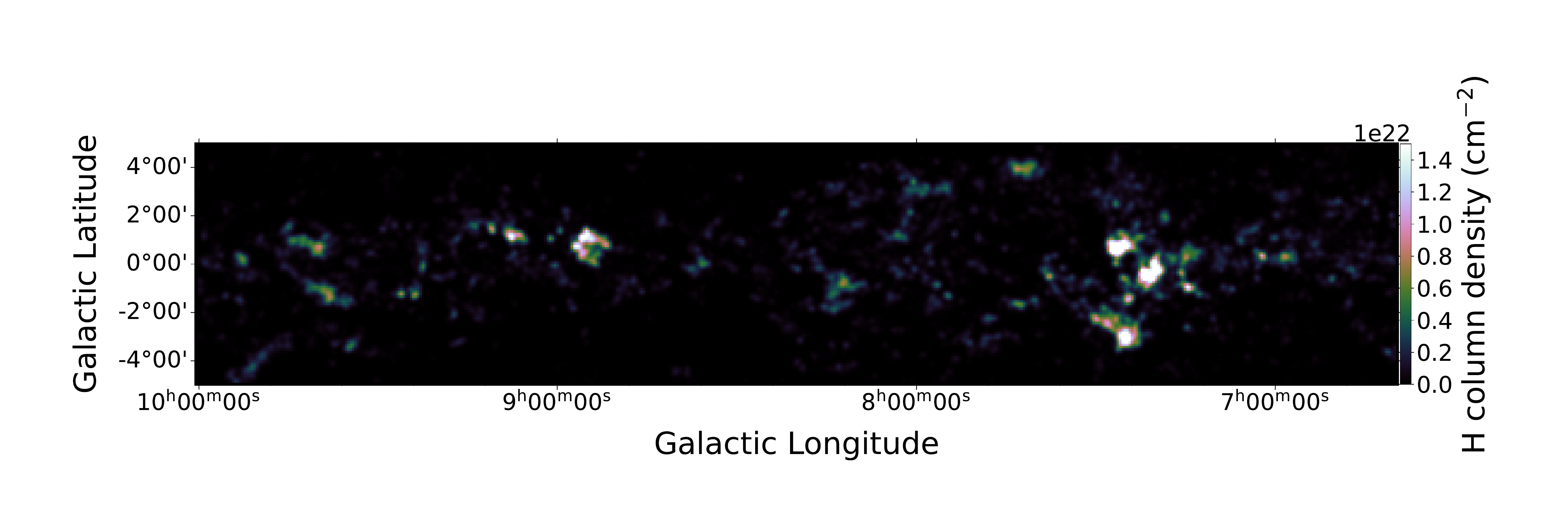}
        \caption{Perseus arm H$_{2}$}
        \label{fig:per_co}
    \end{subfigure}
    \hfill
    \begin{subfigure}[t]{0.48\linewidth}
        \centering
        \includegraphics[width=\linewidth]{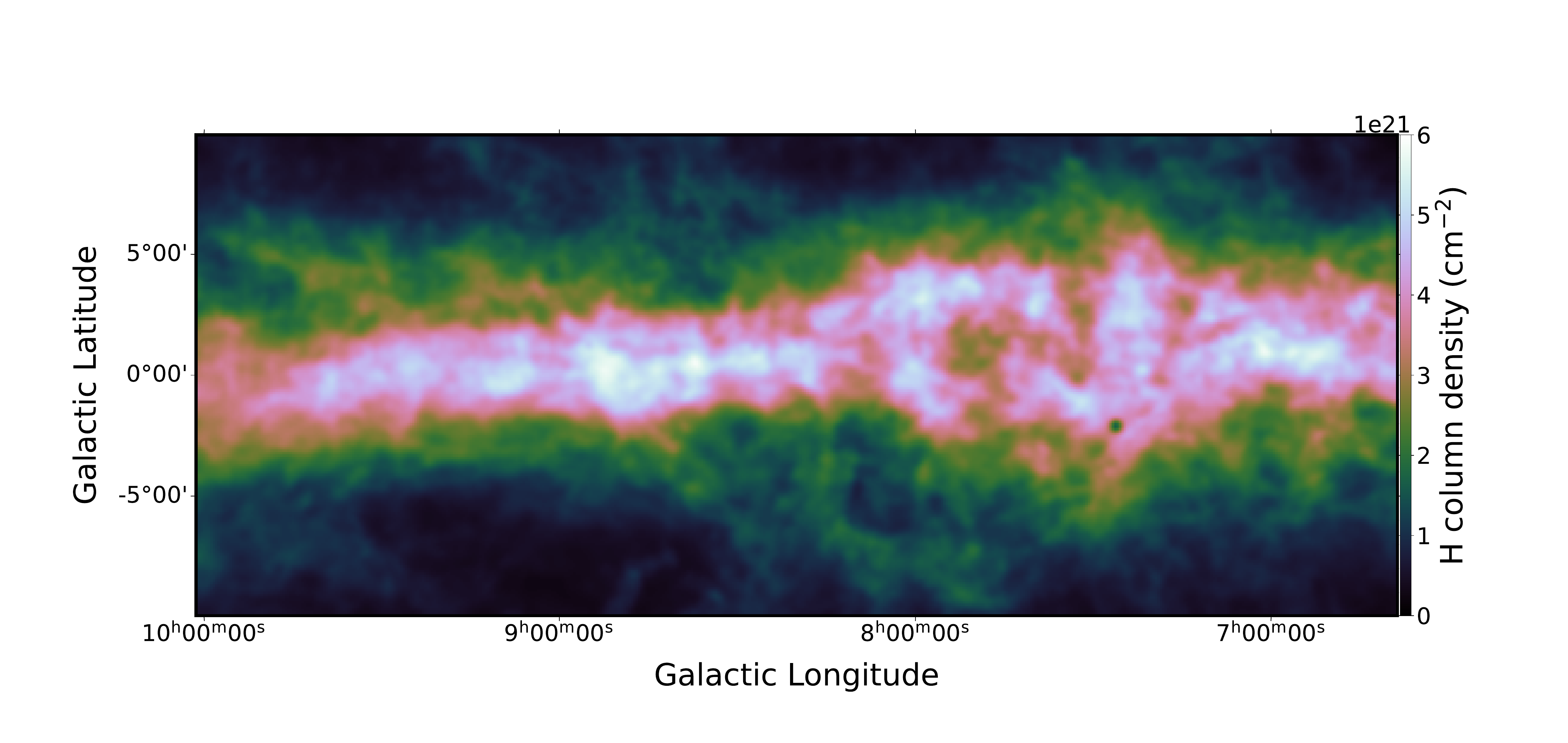}
        \caption{Perseus arm \HI}
        \label{fig:per_hi}
    \end{subfigure}
    
    \vspace{0.3\baselineskip}
    
    % 第三行：Out arm
    \begin{subfigure}[t]{0.48\linewidth}
        \centering
        \includegraphics[width=\linewidth]{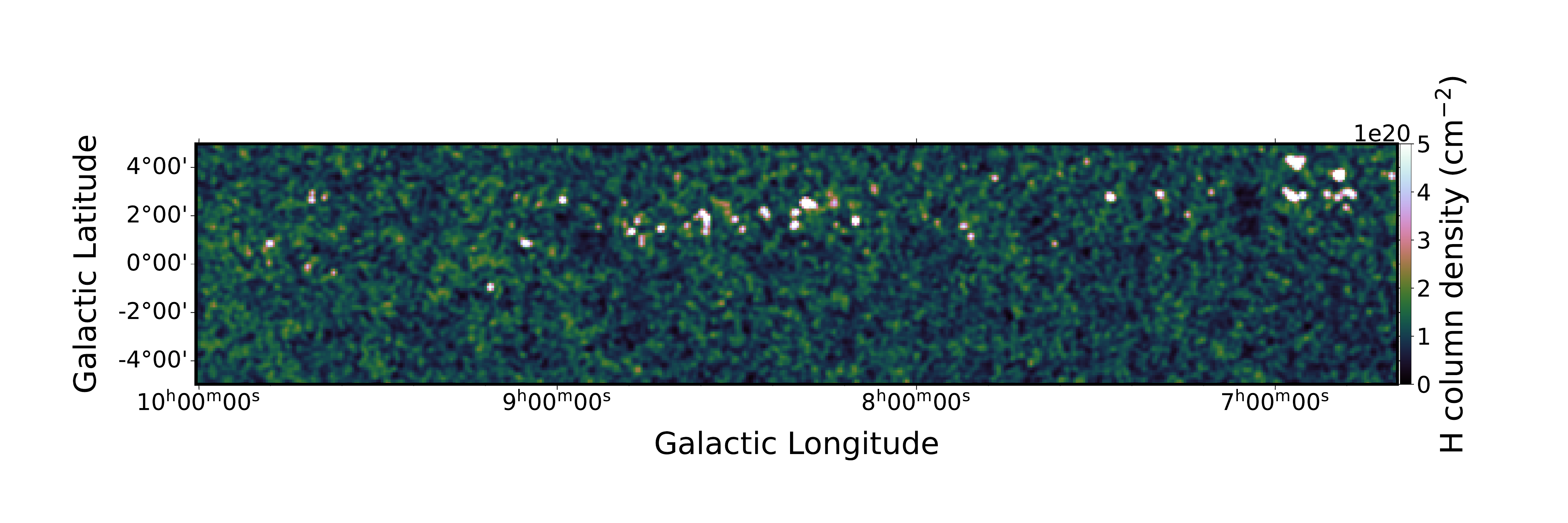}
        \caption{Outer arm H$_{2}$}
        \label{fig:out_co}
    \end{subfigure}
    \hfill
    \begin{subfigure}[t]{0.48\linewidth}
        \centering
        \includegraphics[width=\linewidth]{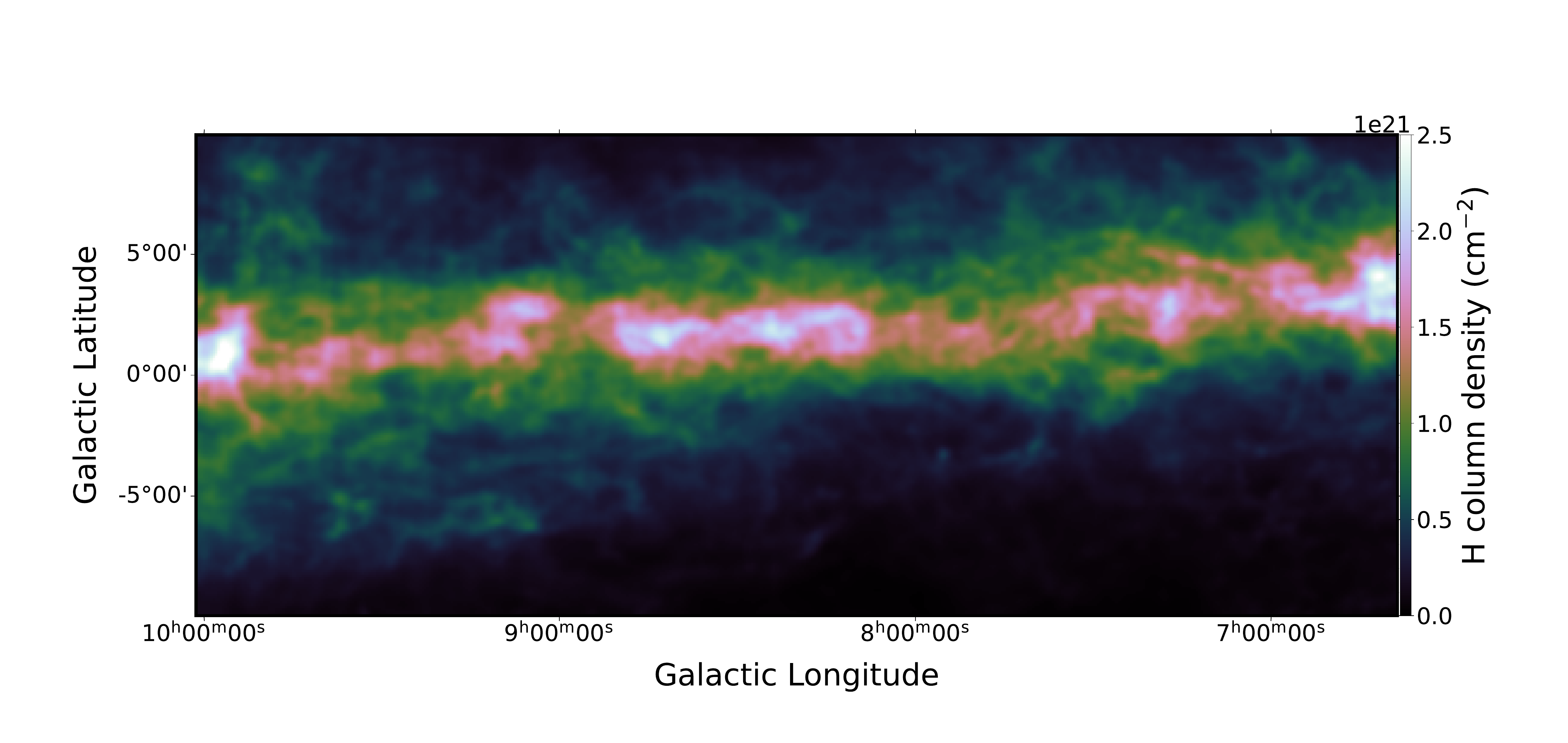}
        \caption{Outer arm \HI}
        \label{fig:out_hi}
    \end{subfigure}
    
    \vspace{0.3\baselineskip}
    
    % 第四行：Dust & DNM
    \begin{subfigure}[t]{0.48\linewidth}
        \centering
        \includegraphics[width=\linewidth]{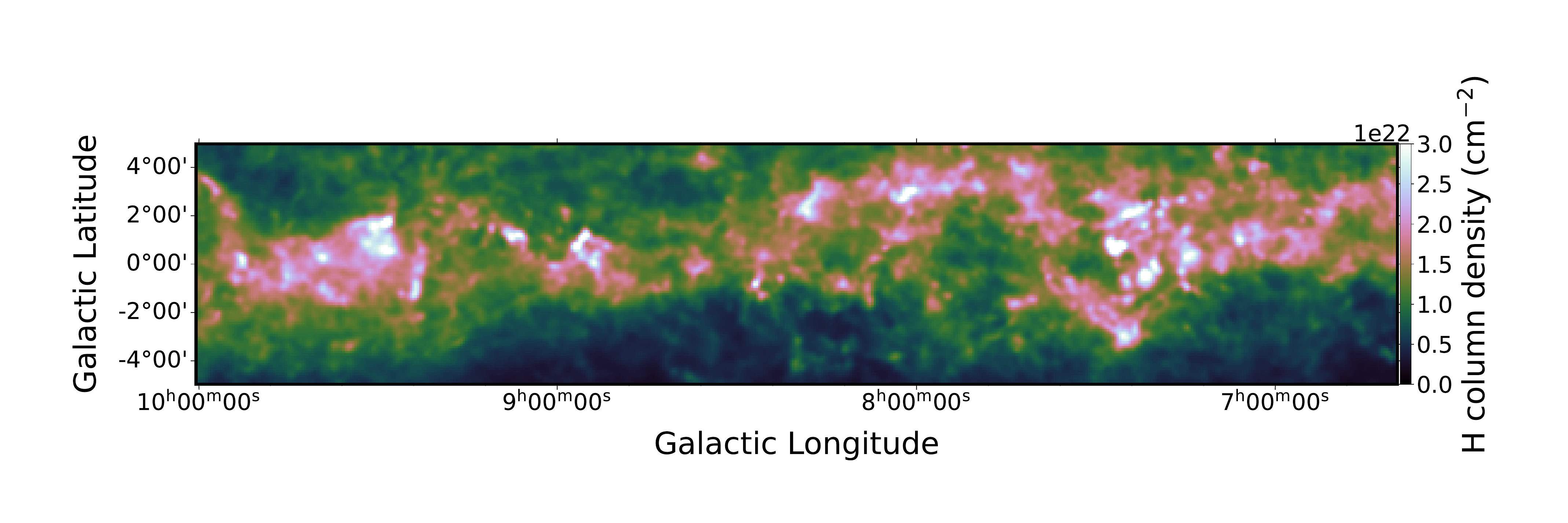}
        \caption{Dust}
        \label{fig:dust}
    \end{subfigure}
    \hfill
    \begin{subfigure}[t]{0.48\linewidth}
        \centering
        \includegraphics[width=\linewidth]{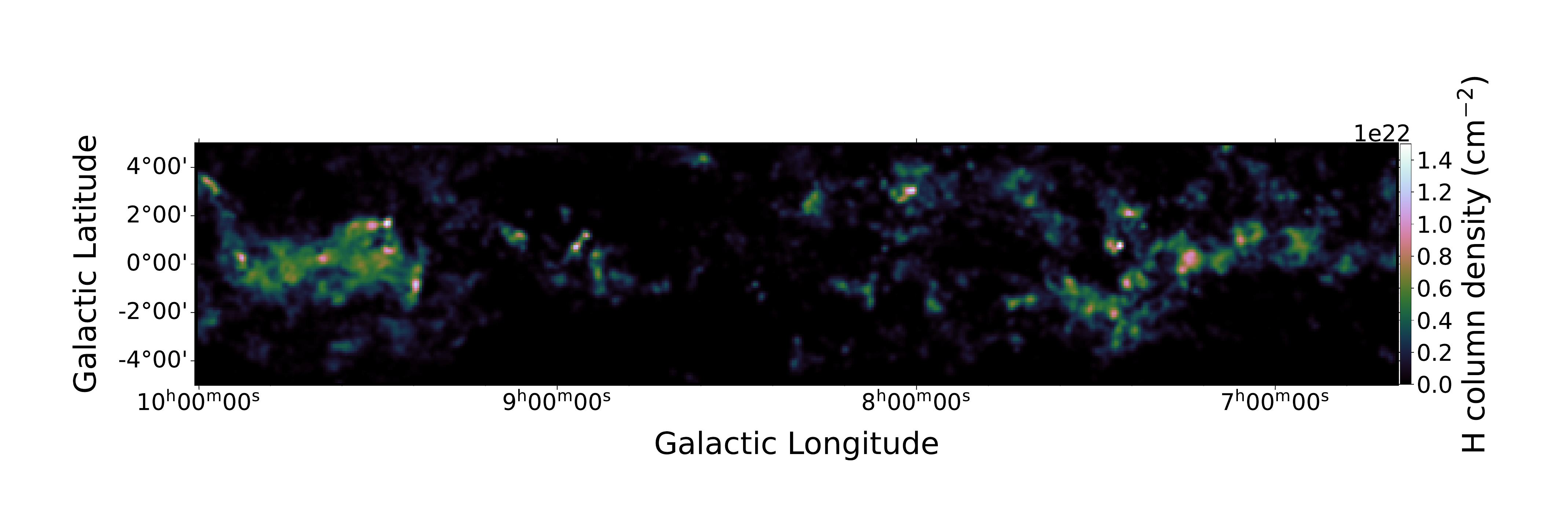}
        \caption{DNM}
        \label{fig:DNM}
    \end{subfigure}
    
    \caption{ Gas templates of the Local, Perseus, and Outer arms (panels a–f), dust template (panel g), and DNM template (panel h), as generated in APPENDIX~\ref{sec:gas}. All maps are scaled to an equivalent total hydrogen column density for visualization purposes, allowing a direct comparison between different gas tracers. The H I component is directly converted to hydrogen column density. The CO component is converted to H$2$ column density using $X_{\rm CO}$=$\rm 2.0 \times 10^{20}\ cm^{-2}\ K^{-1}\ km^{-1}\ s$ and then expressed as total hydrogen column density accounting for the two hydrogen atoms in each H$_2$ molecule. The dust and DNM templates are converted to equivalent hydrogen column density by calibrating the dust opacity against the \HI\-derived gas column density. This scaling is used only for display; the likelihood fit is performed using the original templates, and therefore the fitted normalization parameters are unaffected.}
    \label{fig:all_templates}
\end{figure*}

We investigated three different gas phases, i.e., the H$_{2}$, the \HI, and the DNM. First, we used the CO data from DR1 of the MWISP survey\cite{MWISP_DR1}  to trace the $\rm H_{2}$. We took $N({\rm H_{2}}) = X_{\rm CO} \times W_{\rm CO}$ \citep{Lebrun1983}, where 
$X_{\rm CO}$ is the $\rm H_{2} / CO$ conversion factor chosen to be $\rm 2.0 \times 10^{20}\ cm^{-2}\ K^{-1}\ km^{-1}\ s$ as suggested by \cite{Dame01} and \cite{Bolatto13}. Taking into account that this is an average value within the Galaxy, we attempted to correct it using \gray data, and the results can be seen in Sec.~\ref{sec:data}. However, as a spatial distribution model, changes in normalization do not affect the fitting. Therefore, we continue to use the aforementioned $X_{\rm CO}$ value here for the time being. The obtained templates (scaled to the H column density) of different spiral arms can be seen in Fig.~\ref{fig:loc_co}, Fig.~\ref{fig:per_co}, and Fig.~\ref{fig:out_co}. We refer to them as the H$_{2}$ templates.

Then for the \HI\ density, we used the data cube of HI4PI, which is a 21-cm all-sky database of Galactic \HI\ \citep{HI4PI16}. Under an optically thin assumption, the \HI\ column density can be estimated as:

\begin{equation}
N_{\rm HI} = -1.83 \times 10^{18}~\mathrm{cm}^{-2}\int \left( \frac{\mathrm{d}v}{1~\mathrm{km}~\mathrm{s}^{-1}}\right) \left(\frac{T_{\rm B}}{1~\mathrm{K}}\right),
\end{equation}

Where $T_{\rm B}$ is the brightness temperature of the \HI\ emission. 

We selected a region spanning from $l=100\deg$ to 150$\deg$ and $b=-10\deg$ to 10$\deg$ as the spatial range for the gas templates. However, due to the lack of CO observational data in areas where $|b| > 5\deg$ and the predominance of \HI\ in those regions, the $\rm H_{2}$ template constructed from the CO data was derived only from the latitude range of $-5\deg$ to 5$\deg$.

The velocity ranges of different spiral arms vary with changes in Galactic longitude. Based on the results from Figure 5 of \cite{du2016}, we integrated the data over the appropriate velocity intervals and ultimately derived the column density distributions of $\rm H_{2}$ and \HI\ for each spiral arm. Since the new spiral arm was not significantly observed in the data from DR1 of the MWISP survey \citep{MWISP_DR1} used in our study, we did not differentiate between the Outer arm and the new arm in the integration. In the following sections, this portion of the gas is collectively referred to as the "Outer arm". And for the Local arm, we integrated the gas with a LSR velocity of more than $-20 \rm~km/s$. The obtained templates (scaled to the H column density) of different spiral arms can be seen in Fig.~\ref{fig:loc_hi}, Fig.~\ref{fig:per_hi}, and Fig.~\ref{fig:out_hi}.

Following the characterization of the gas phases \HI\ and $\rm H_{2}$, these traditional tracers do not account for a significant component of the interstellar medium. This component, known as the DNM, comprises gas predominantly in the form of cold atomic hydrogen and diffuse molecular hydrogen that is not adequately traced by either \HI\ 21 cm line emission or $^{12}$CO line emission. The DNM exists at the interface between the atomic and molecular phases, where the 21 cm line may be optically thick and CO molecules may be absent or under-excited. The primary method for tracing the large-scale distribution of the DNM is through interstellar dust emission. This approach has become standard practice in Fermi-LAT studies of the interstellar medium. In particular, \citet{Acero2016} derived DNM templates from the positive residuals obtained after fitting dust tracers with \HI and $\rm H_{2}$ gas maps, and similar methodologies have been adopted in subsequent investigations of nearby molecular clouds and diffuse interstellar gas \citep[e.g.,][]{Remy2017,Youssef2024}. We utilize an all-sky map of dust opacity from the Planck collaboration \citep{2014Planck}. The underlying assumption is that for a uniform dust-to-gas ratio and uniform dust grain properties, the dust opacity map serves as a proxy for the total gas column density. A multi-step linear regression is then performed to fit the dust reddening map as a linear combination of the \HI\ and $\rm H_{2}$ column density maps integrated over the entire velocity range. The DNM template is derived from the positive residuals of the linear fit described above. These residuals represent excess dust emission that is not accounted for by the \HI\ and CO maps, which we attribute to the presence of additional gas in the DNM. 
The linear fitting procedure also produces negative residuals, indicating regions where the combined \HI\ and CO model overpredicts the observed dust opacity. These deficits are interpreted not as part of the DNM but rather as potential systematic effects. We note that the assumption of a uniform dust-to-gas ratio is only approximate. Spatial variations in dust properties may introduce systematic uncertainties in the derived dust template and DNM template. Nevertheless, dust opacity remains a useful large-scale tracer of the total gas column density and has been widely adopted in previous studies \citep[e.g.,][]{Remy2017,Youssef2024}.

%% For this sample we use BibTeX plus aasjournalv7.bst to generate the

\bibliography{apj}{}
\bibliographystyle{aasjournalv7}



\end{document}